\documentclass[
    aps,12pt,final,notitlepage,
    oneside,onecolumn,nofootinbib,
    superscriptaddress,noshowpacs,
    nolongbibliography,centertags
]{revtex4-2}

\usepackage[english]{babel}

\usepackage[usenames,dvipsnames]{xcolor} % link colors tweak
\usepackage[
    colorlinks = true,
    urlcolor = MidnightBlue,
    linkcolor = Red,
    citecolor = NavyBlue
]{hyperref}

\usepackage{graphicx}
\graphicspath{{figures/}}

\usepackage{amsmath,amssymb}

\usepackage{hyperref}
\hypersetup{
    pdfauthor = {A. V. Glushkov, K. G. Lebedev, A. Saburov},
    pdftitle = {Muons in EASs with $E_0 = 10^{19}$~eV according to data of the Yakutsk Array},
    pdfsubject = {extensive air showers},
    pdfkeywords = {ultra-high energy cosmic rays, extensive air showers, muon puzzle, the Yakutsk array}
}

\newcommand{\DOI}[2] {%
    \href{https://doi.org/#1}{#2}%
}
\newcommand{\arXiv}[2] {%
    arXiv:~\href{https://arXiv.org/abs/#1}{#1} [#2]%
}

\newcommand{\eq}[1]{(\ref{#1})}

\newcommand{\Emu}{E_{\mu}} % threshold energy
\newcommand{\E}{E_0}

\newcommand{\RhoMu}{\rho_{\mu}}
\newcommand{\RhoMuExp}{\RhoMu^{\text{exp}}}
\newcommand{\RhoMuP}{\RhoMu^p}
\newcommand{\RhoMuFe}{\RhoMu^{\text{Fe}}}
\newcommand{\RhoMuSOO}{\rho_{\mu,600}}
\newcommand{\bmu}{b_{\mu}}

\newcommand{\RhoS}{\rho_{\text{s}}}
\newcommand{\RhoSSOO}{\rho_{\text{s},600}}
\newcommand{\bs}{b_{\text{s}}}

\newcommand{\meancos}{\left<\cos\theta\right>}

\newcommand{\Ethin}{E_{\text{thin}}}
\newcommand{\wmax}{w_{\text{max}}}

\newcommand{\qgs}{{\sc qgsj}et01}
\newcommand{\qgsii}{{\sc qgsj}et-{\sc ii}.04}

\newcommand{\eposlhc}{{\sc epos-lhc}}

\newcommand{\fluka}{{\sc fluka}2011}
\newcommand{\corsika}{{\sc corsika}}

\newcommand{\degr}{^{\circ}}       % degrees

\begin{document}

\title{Muons in EASs with $\E = 10^{19}$~eV according to data of the Yakutsk Array}

\author{A.\,V.\,Glushkov}
\email{glushkov@ikfia.ysn.ru}

\author{K.\,G.\,Lebedev}
\email{LebedevKG@ikfia.ysn.ru}
\affiliation{Yu.G.Shafer Institute of cosmophysical reserach and aeronomy of Siberian branch of the Russian Academy of Sciences,\\
31 Lenin ave., Yakutsk, 677027, Russia}

\author{A.\,V.\,Saburov}
\email{vs.tema@gmail.com}

%\dates{\today}{*}

\begin{abstract}
    Lateral distribution functions of particles in extensive air showers with the energy $\E \simeq 10^{19}$~eV recorded by ground-based and underground scintillation detectors with a threshold of $\Emu \simeq 1.0 \times \sec\theta$~GeV at the Yakutsk array during the continuous observations from 1986 to 2016 have been analyzed using events with zenith angles $\theta \le 60\degr$ functions have been compared to the predictions obtained with the \qgs{} hadron interaction model by applying the \corsika{} code. The entire dataset indicates that cosmic rays consist predominantly of protons.
\end{abstract}

\maketitle

\section{Introduction}

In recent years, a problem of muon excess in extensive air showers (EAS) has
arisen in several experiments in comparison with model
predictions~\cite{bib:1}. Many collaborations are involved in solving this problem. Different datasets are compared using the parameter

\begin{equation}
    z = \frac{
        \ln{\RhoMuExp} - \ln{\RhoMuP}
    }{
        \ln{\RhoMuFe} - \ln{\RhoMuP}
    }\text{,}
    \label{eq:1}
\end{equation}

\noindent
where $\RhoMuExp$ is the muon density measured in the experiment and $\RhoMuP$ and $\RhoMuFe$ are the muon densities calculated for EASs initiated by primary protons and iron nuclei in this experiment, respectively. A combined analysis of the data from eight research groups (EAS-MSU, Ice-Cube Neutrino Observatory, KASCADE-Grande, NEVOD-DECOR, Pierre Auger Observatory, SUGAR, Telescope Array, and Yakutsk) showed that model calculations are in agreement with muon measurements to $10^{16}$~eV. However, the situation changes with a further increase in the primary energy. A significant spread of the $z$ value is observed, especially in inclined showers~\cite{bib:2} and at large distances from the shower axis~\cite{bib:3}. Muon densities measured at the Yakutsk array in showers with $\E \ge 10^{18}$~eV and $\meancos = 0.9$ at a distance of 300~m from the axis gave the value $z \simeq 0$ with the \qgs{} model and negative values with the \qgsii{} and \eposlhc{} models~\cite{bib:1}. In~\cite{bib:4}, the fraction of muons was investigated at distances 300, 600, and 1000~m from the axis in showers with $\E \simeq 10^{17.7 - 19.5}$~eV and $\meancos = 0.9$. The agreements with the \qgs{} model was confirmed for primary protons ($z \simeq 0$). Here, we continue to study the fraction of muons in EASs with the energy $\simeq 10^{19}$~eV in a wide range of zenith angles.

\section{Lateral distribution of EAS particles}

\subsection{Calculation of Mean Lateral Distribution Functions}

\begin{figure}[!htb]
    \centering
    \includegraphics[width=0.65\textwidth]{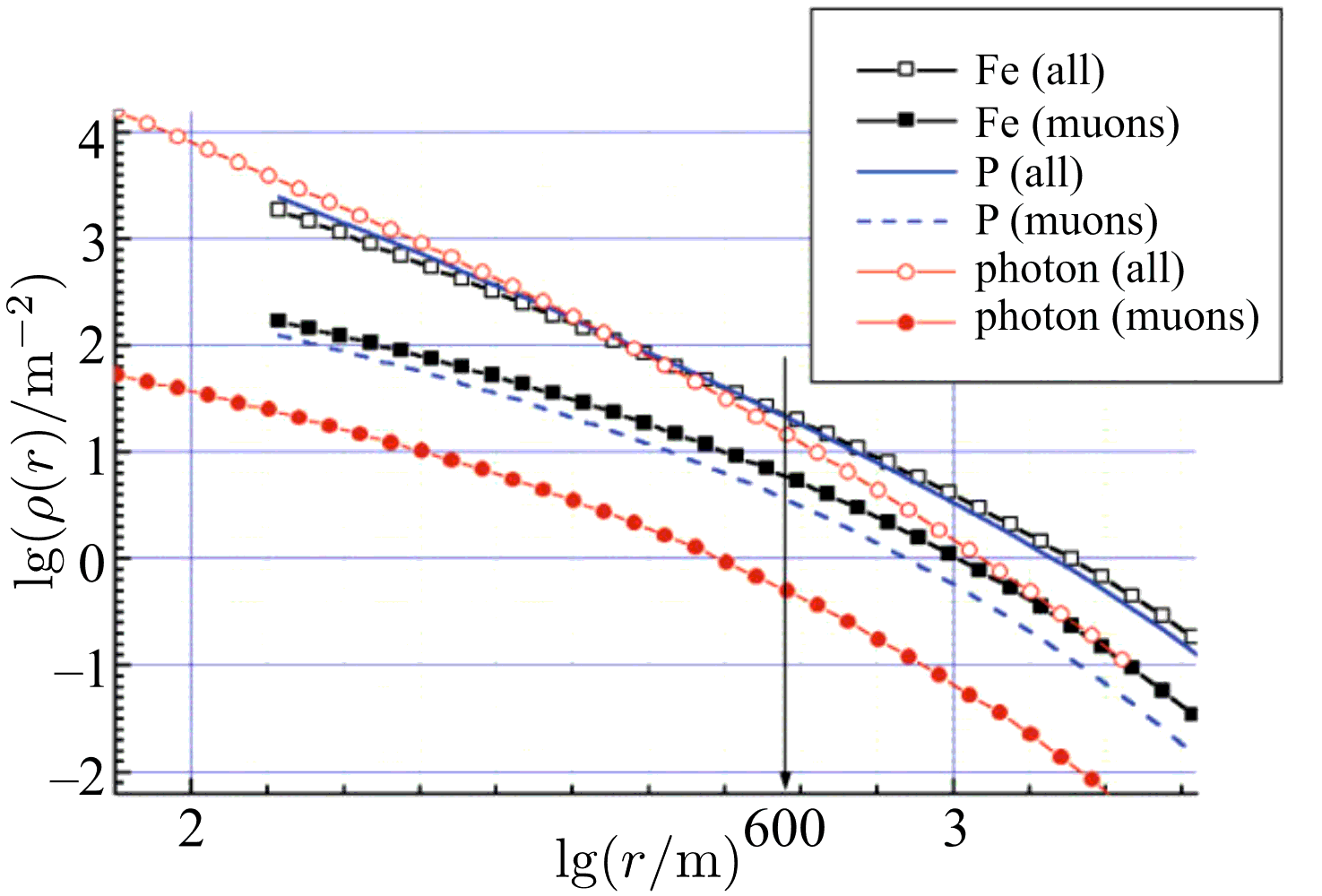}
    \caption{Lateral distribution functions of responses in ground-based and underground scintillation detectors with a threshold of $\Emu \simeq 1.0 \times \sec\theta$~GeV in EASs initiates by primary particles with the energy $\E \simeq 10^{19}$~eV and $\cos\theta = 0.90$ obtained with the \qgs{} model~\cite{bib:6}.}
    \label{fig:1}
\end{figure}

\begin{figure}[!htb]
    \centering
    \includegraphics[width=0.65\textwidth]{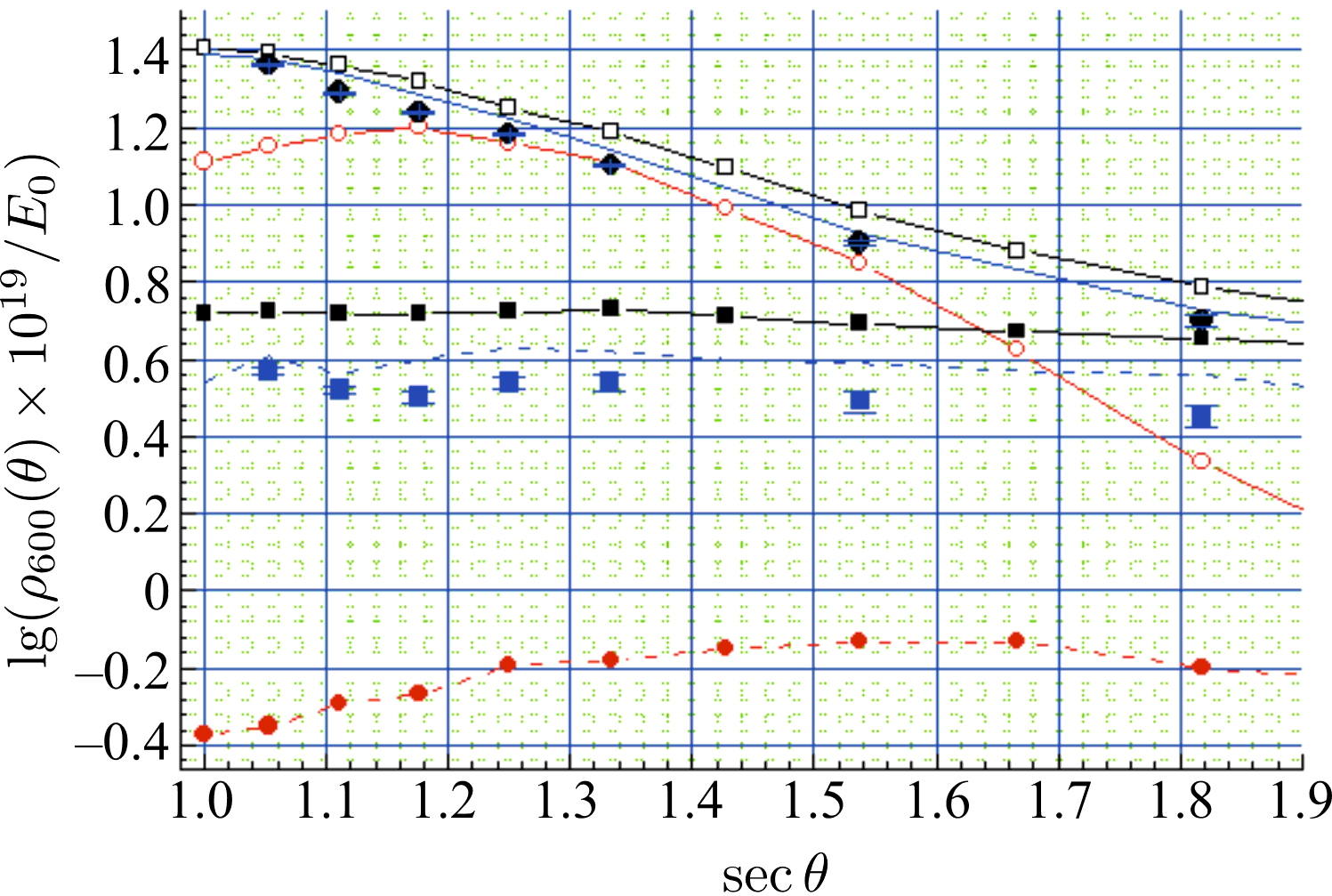}
    \caption{Zenith-angular dependences of responses in ground-based and underground scintillation detectors in showers initiated by primary particles with the energy $\E \simeq 10^{19}$~eV at a distance 600~m from the axis obtained with the \qgs{} model~\cite{bib:6} (notation is the same as in Fig.~\ref{fig:1}). Dark circles and squares are experimental data (see the main text).}
    \label{fig:2}
\end{figure}

The responses of ground-based and underground scintillation detectors of the Yakutsk array to EASs initiated by primary particles with the energy above $10^{17}$~eV were calculated in\cite{bib:5, bib:6} with a set of artificial showers generated with the \corsika{} code~\cite{bib:7} using the \qgs~\cite{bib:8} and \qgsii{} models~\cite{bib:9}. The \fluka{} code~\cite{bib:10} was chosen to describe hadron interactions at energies below 80~GeV. Showers were simulated with zenith angles $0\degr-60\degr$ in the energy range of $10^{17}-10^{19.5}$~eV with a logarithmic step of $\Delta\lg(\E/\text{eV}) = 0.5$. The calculations involved the thin-sampling mechanism~\cite{bib:11} with the thinning level $\Ethin = 10^{-6}-10^{-5}$ and the weight limit for all components $\wmax = \E \cdot \Ethin$. For each set of the input parameters $(\E, \theta)$, from 200 to 500 events were generated. Using these events, mean lateral distribution functions (LDFs) of the detector response were obtained with the radial binning of the distance from the axis with a $\Delta\lg(r/\text{m}) = 0.04$ step. Figure~\ref{fig:1} exemplifies a calculated LDF for the response in ground-based (all) and underground (muons) scintillation detectors of the Yakutsk array in events originated from different primary particles. Figure~\ref{fig:2} shows responses from particles in EASs with $\E = 10^{19}$~eV and different zenith angles at a distance of 600~m from the axis. All densities were converted to $\E = 10^{19}$~eV by multiplication by normalization factors $10^{19} / \left<\E\right>$.

\subsection{Event Selection and Processing}

The average densities $\left<\RhoS(\theta)\right>$ and $\left<\RhoMu(\theta)\right>$ of all particles and muons in EASs with the threshold energy $\Emu \simeq 1.0 \times \sec\theta$~GeV were considered obtained at a distance of 600~m from the axis in events with mean arrival zenith angles $\meancos = 0.95, 0.90, 0.85, 0.80, 0.75, 0.65$, and $0.55$. Experimental LDFs of both components were calculated in zenith-angle intervals $\Delta\cos\theta = 0.1$ with the energy increment $\Delta\lg(\E/\text{eV}) = 0.2$. We selected showers whose axes were located in a circle with a radius of 1~km around the array center and were determined with an accuracy of no worse than 50~m. The accuracy of evaluation of $\RhoSSOO(\theta)$ in individual events was above 10\%. The primary energy of showers was determined by the formula~\cite{bib:12}

\begin{equation}
    \E = (3.76 \pm 0.3) \times
    10^{17}(\RhoSSOO(0\degr)^{1.02 \pm 0.02}~\text{[eV],}
    \label{eq:2}
\end{equation}

\noindent
where

\begin{equation}
    \RhoSSOO(0\degr) = \RhoSSOO(\theta) \cdot
    \exp\left(
        \frac{(\sec\theta - 1) \times 1020}{\lambda}
    \right)~\text{[m$^{-2}$].}
    \label{eq:3}
\end{equation}

\begin{figure}
    \centering
    \includegraphics[width=0.65\textwidth]{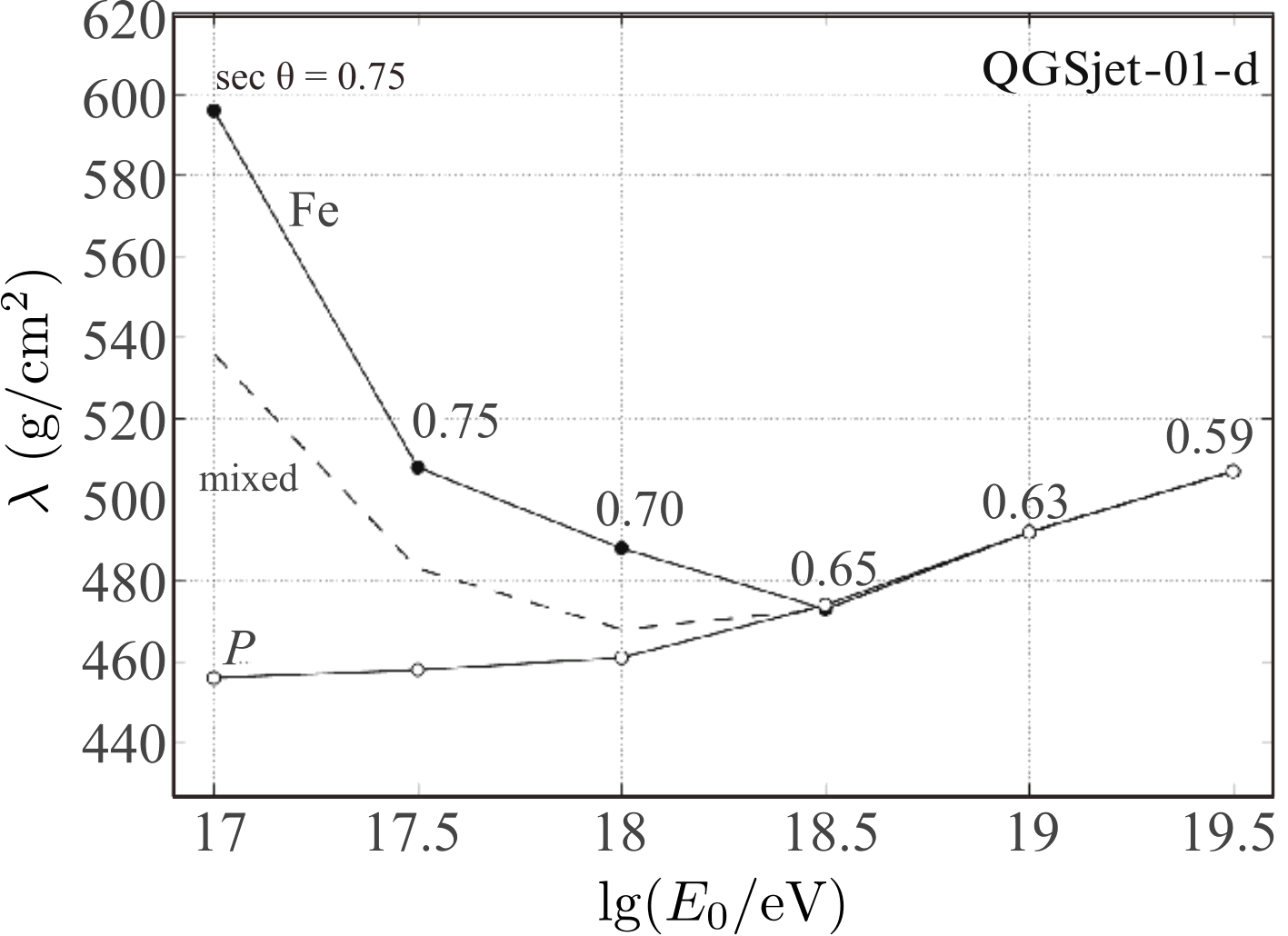}
    \caption{Energy dependence of the absorption range in \eq{eq:3} used in the recalculation of $\RhoSSOO(\theta)$ from inclined to vertical showers according to the \qgs{} model for primary protons ($p$), mixed composition (mixed), and iron nuclei (Fe). Numbers next to the data points indicate limit values of $\cos\theta$~\cite{bib:12}.}
    \label{fig:3}
\end{figure}

\noindent
Here, $\lambda$ is the absorption length shown in Fig.~\ref{fig:3} and $\RhoSSOO(\theta)$ is the density determined in the experiment. The mixed composition was taken from the experiment reported in~\cite{bib:12}. Relation \eq{eq:2} unambiguously relates $\RhoSSOO(0\degr)$ to $\E$ at any cosmic ray composition since LDFs of cascade particles in EASs intersect each other at $r \simeq 600$~m. In the case of primary photons, all three LDFs intersect at $r \simeq 450$~m (Fig.~\ref{fig:1}). When calculating LDFs, particle densities in individual events were multiplied by a normalization coefficient $\left<\E\right>/\E$ (where $\left<\E\right>$ is the average energy in a group) and averaged in distance intervals $[\lg(r_i), \lg(r_i) + 0.04]$. Average particle densities in these intervals were determined as:

\begin{equation}
    \left<\RhoS(r_i)\right> = 
    \frac{1}{N}
    \sum_{k = 1}^{N} \rho_k(r_i)\text{,}
    \label{eq:4}
\end{equation}

\noindent
where $N$ is the number of detector readings at a given distance from the axis. The resulting mean LDFs were approximated by the function

\begin{equation}
    \RhoS(r,\theta) = \RhoSSOO(\theta) \cdot
    \left(
        \frac{600}{r}
    \right)^2 \cdot
    \left(
        \frac{608}{r + 8}
    \right)^{\bs - 2} \cdot
    \left(
        \frac{600 + r_1}{r + r_1}
    \right)^{10}\text{,}
    \label{eq:5}
\end{equation}

\noindent
where $r_1 = 10^4$~m and $\RhoSSOO(\theta)$ and $\bs$ are the free parameters determined by minimizing $\chi^2$.

The muon LDF was constructed in a similar way. Average particle densities were determined as

\begin{equation}
    \left<\RhoMu(r_i)\right> = 
    \frac{1}{N_1 + N_0}
    \sum_{n = 1}^{n_1}
    \RhoMu(r_i)\text{,}
    \label{eq:6}
\end{equation}

\noindent
where $N_1$ and $N_0$ are the numbers of nonzero and zero readings of muon detectors, respectively, at distances in the intervals $[\lg(r_i), \lg(r_i) + 0.04]$.

Zero readings $N_0$ correspond to cases where detectors do not record a single muon while being in the accepting mode. The LDF was approximated by the function

\begin{equation}
    \RhoMu(r,\theta) = \RhoMuSOO(\theta) \cdot
    \left(
        \frac{600}{r}
    \right)^{0.75} \cdot
    \left(
        \frac{880}{r + 280}
    \right)^{\bmu - 0.75} \cdot
    \left(
        \frac{600 + r_1}{r + r_1}
    \right)^{6.5}\text{,}
    \label{eq:7}
\end{equation}

\noindent
where $r_1 = 2000$~m and $\bmu$ and $\RhoMuSOO(\theta)$ are the free parameters determined by minimizing $\chi^2$.

\section{Results and discussion}

\begin{figure}
    \centering
    \includegraphics[width=0.65\textwidth]{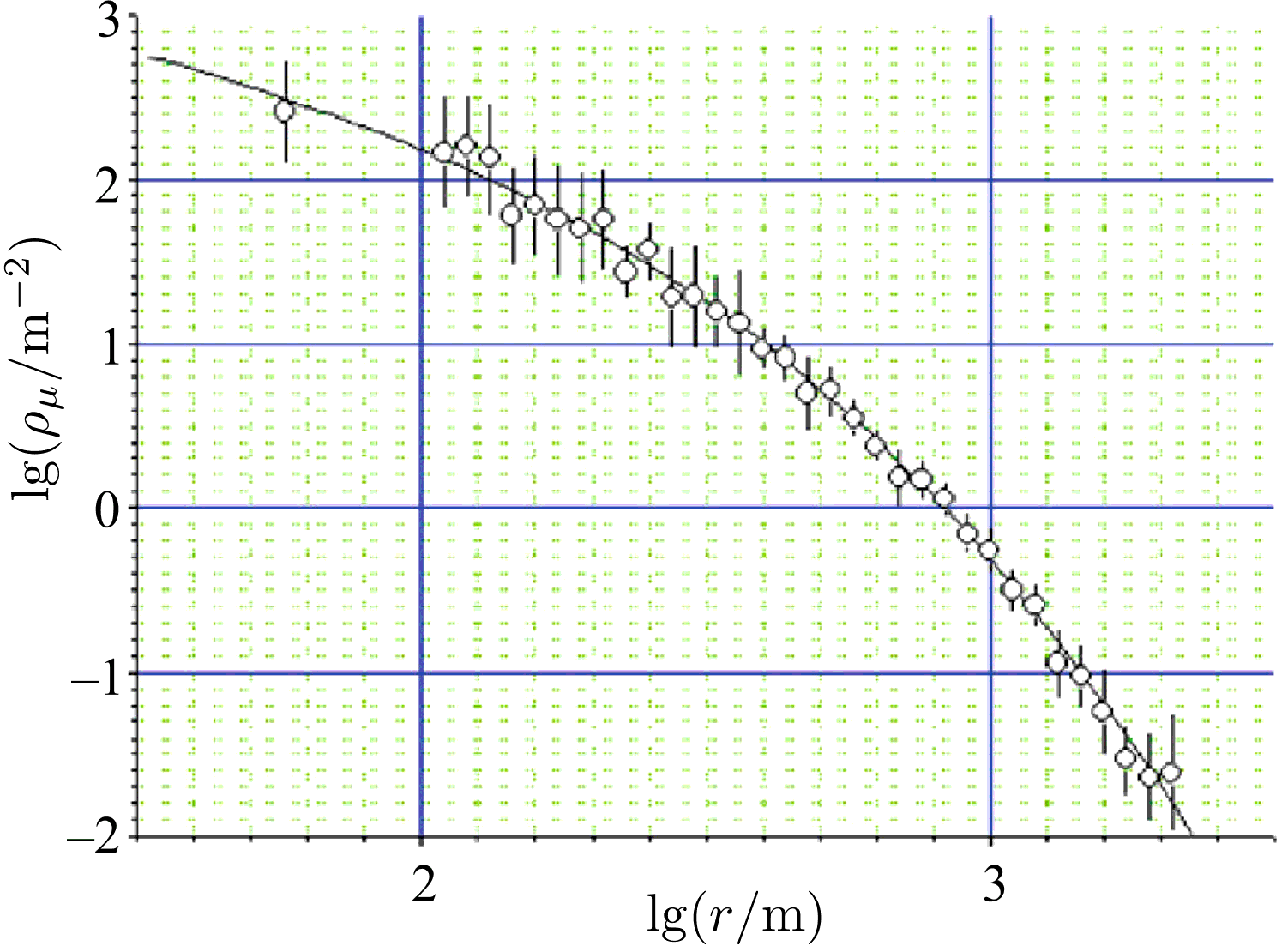}
    \caption{Mean lateral distribution function of muons in showers with $\left<\E\right> = 8.93 \times 10^{18}$~eV and $\meancos \simeq 0.8$. The line is approximation \eq{eq:7} with the parameters $\left<\bmu\right> = 2.140 \pm 0.027$ and $\left<\lg(\RhoMu)\right> = 0.488 \pm 0.02$. The processing agreement criterion for all data points is $\chi^2 = 25.1$.}
    \label{fig:4}
\end{figure}

Figure~\ref{fig:4} presents one of the mean muon LDFs derived from the experimental data. The density at $r = 600$~m divided by the average primary energy of EASs equals $\lg(\RhoMuSOO(37\degr) \times 10^{19} / \left<\E\right>) = 0.538 \pm 0.017$. Other experimental data were obtained in a similar manner.

It is seen in Fig.~\ref{fig:2} that responses in ground-based and underground detectors from EAS particles are lower than expected responses from primary protons and muon densities are significantly lower than expected values. There are several possible reasons for this result. One of them is the energy estimation in the experiment. The first term in Eq. \eq{eq:2} reflects a systematic error of 8\% from the uncertainty of the array calibration method itself~\cite{bib:12}. Equation \eq{eq:3} introduces from 0 to 15\% due to zenith-angle uncertainty during the transition from $\RhoSSOO(\theta)$ to the vertical direction. This error is due to dependences of the parameter $\lambda$ on the primary nucleus atomic number in Eq. \eq{eq:3} (see, e.g., Fig.~\ref{fig:2}) and on the hadron interaction model. Neither of them are known a priori.

\begin{table}[!htb]
    \centering
    \caption{Values of $z$ parameter \eq{eq:1} in groups of showers with
different zenith angles.}
    \label{t:1}
    \begin{tabular}{%
        p{0.1\textwidth}p{0.1\textwidth}%
        p{0.1\textwidth}p{0.1\textwidth}%
        p{0.1\textwidth}p{0.1\textwidth}%
        p{0.1\textwidth}p{0.1\textwidth}%
    }
        \hline
        $\sec\theta$ & 1.052 & 1.111 & 1.176 & 1.250 & 1.333 & 1.538 & 1.818 \\
        \hline
        $z$ & 0.0 & 0.0 & -0.4 & -0.7 & -0.7 & -0.7 & -0.7 \\
        $\pm\Delta z$ & 0.10 & 0.10 & 0.2 & 0.2 & 0.2 & 0.3 & 0.3 \\
        \hline
    \end{tabular}
\end{table}

To understand the above result, we assume that shower energy was overestimated by the difference between theory and experiment in the case of ground- based detectors, i.e., by 10\%. If the shower energy is reduced by this value, densities measured by ground-based detectors and presented by filled circles in Fig.~\ref{fig:2} will be in agreement with the simulated values. Muon densities shown in Fig.~\ref{fig:2} will also increase by 10\% after energy re-evaluation. The first two sets of the data (at $\sec\theta \simeq 1.05$ and $1.11$) will agree with the \qgs{} model, whereas the other will remain 10\% below the expected values. The results of applying Eq. \eq{eq:1} to the experimental data are summarized in Table~\ref{t:1}. The indicated errors include both statistical errors in the calculation of mean LDFs and errors in the reconstruction of individual events (arrival direction, axis coordinates, and energy estimation). It is difficult to distinguish between them and this is not necessary. They are accumulated in the average values $\left<\RhoSSOO(\theta)\right>$ and $\left<\RhoMuSOO(\theta)\right>$ (see, e.g., Fig.~\ref{fig:3}). Values presented in first two columns of Table~\ref{t:1} agree with our estimates given in~\cite{bib:1}. The other data are physically meaningless in the z parameter and cannot be explained by methodical distortions of the experiment. We observe another muon puzzle in inclined showers with the energy $\simeq 10^{19}$~eV, but with the opposite sign: a muon deficit is observed in measured densities in comparison with the \qgs{} and \qgsii{} models for primary protons. All above speculations on a hypothetical 10\% shift of the energy are conditioned only by the uncertainty in the first term of Eq. \eq{eq:2}, which is due to the technique of absolute primary energy calibration at the Yakutsk array~\cite{bib:12}. We do not exclude the possibility of its further refinement as the experiment progresses.

\begin{figure}[!htb]
    \centering
    \includegraphics[width=0.65\textwidth]{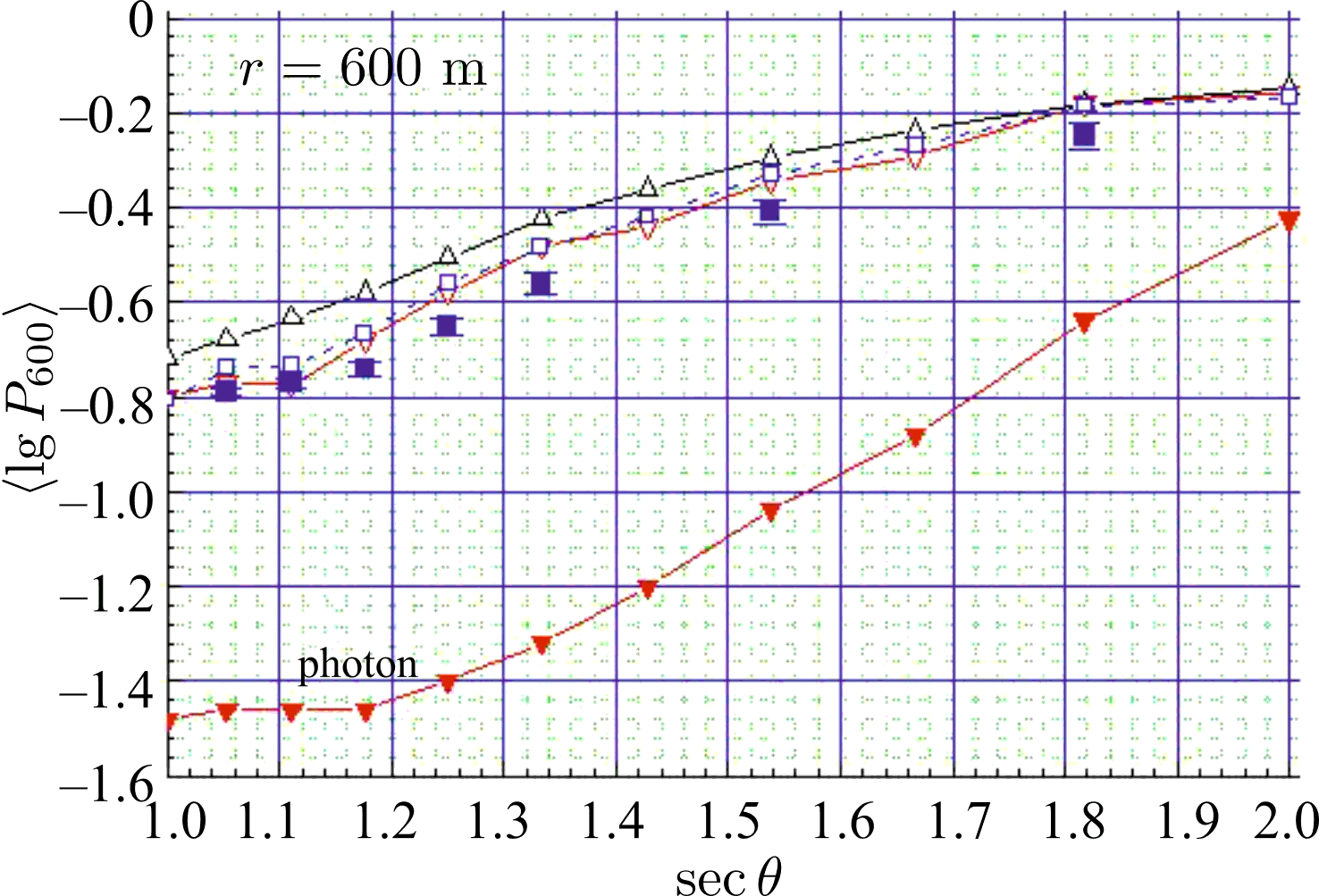}
    \caption{Zenith-angular dependences of the fraction of muons $\left<\lg(\RhoMuSOO /\RhoSSOO)\right>$ at a distance 600~m from the axis in the EAS with $\E = 10^{19}$~eV initiated by (empty upward triangles) primary protons and (empty downward triangles) iron nuclei according to the \qgs{} model and by (empty squares) primary pro- tons according to the \qgsii{} model~\cite{bib:6}. Filled squares are experimental data.}
    \label{fig:5}
\end{figure}

At first glance, the reported results are critically sensitive to experimental errors of energy estimation, but this is not entirely the case. If mean LDFs of both EAS components were obtained from the same source dataset with the energy $\left<\E\right>$, the fraction of muons

\begin{displaymath}
    P_{600} = \frac{\left<\RhoMuSOO\right>}{\left<\RhoSSOO\right>} =
    \frac{
        \left<\RhoMuSOO\right> / \left<\E\right>
    }{
        \left<\RhoSSOO\right> / \left<\E\right>
    }
\end{displaymath}

\noindent
in this set hardly depends on the energy. Figure~\ref{fig:5} shows the fraction of muons obtained from the data presented in Fig.~\ref{fig:2}. The deficit of the measured muon component compared to the \qgs{} and \qgsii{} models for primary protons is directly seen. In our opinion, this problem can be solved under the assumption that the total cosmic ray flux contains a certain fraction of primary photons, which yield almost an order of magnitude less muons (see Figs.~\ref{fig:2} and \ref{fig:3}). There are several estimates of the upper limit of the photon fraction in cosmic rays in this energy region~\cite{bib:13, bib:14, bib:15}. According to the Yakutsk array, it is 10\%~\cite{bib:13}. Among 33 showers with energies above $2 \times 10^{19}$~eV considered in~\cite{bib:12}, there are two events with a low muon content (6\%) with zenith angles of $18\degr$ and $42\degr$. Our calculations showed that the fractions of primary nuclei and photons in the total cosmic ray flux can be estimated by the formulas

\begin{gather}
    w_A = \frac{
        \lg{P_{\text{exp}}(\theta)} - \lg{P_{\gamma}(\theta)}
    }{
        \lg{P_A(\theta)} - \lg{P_{\gamma}(\theta)}
    }\text{,}
    \label{eq:8} \\
    w_{\gamma} = 1 - w_A\text{,}
    \label{eq:9}
\end{gather}

\noindent
respectively. The results are summarized in Table~\ref{t:2}, where only statistical errors following from the analysis of mean LDFs are presented. The first four columns correspond to the proton–photon pair. It is seen that the mean fraction of protons in these groups with $\sec\theta = 1.053$ and $1.111$ (first two rows) is $0.99 \pm 0.01$, which differs from other five rows, where its average value is $0.91 \pm 0.03$. The results for a hypothetical pair iron–photon are also presented in Table~\ref{t:2}. In the latter case, agreement with the experiment can be achieved at the 16\% fraction of primary photons in the total cosmic ray flux.

\begin{table}[!htb]
    \centering
    \caption{Fractions of protons and iron nuclei paired with primary photons in the total cosmic ray flux in showers with different zenith angles.}
    \label{t:2}
    \begin{tabular}{%
        p{0.09\textwidth}%
        p{0.09\textwidth}p{0.09\textwidth}%
        p{0.09\textwidth}p{0.09\textwidth}%
        p{0.09\textwidth}p{0.09\textwidth}%
        p{0.09\textwidth}p{0.09\textwidth}%
    }
    %\begin{tabular}{c|cccc|cccc}
        \hline
        $\sec\theta$ &
            $w_p$              &    $\pm\Delta w_p$             &
            $w_{\gamma}$       &    $\pm wd_{\gamma}$           & 
            $w_{\text{Fe}}$    &    $\pm\Delta w_{\text{Fe}}$   &
            $w_{\gamma}$       &    $\pm wd_{\gamma}$           \\
        \hline
        1.053   & 0.98 & 0.02 & 0.02 & 0.02 & 0.85 & 0.02 & 0.15 & 0.02 \\
        1.111   & 1.00 & 0.02 & 0.00 & 0.02 & 0.83 & 0.02 & 0.17 & 0.02 \\
        1.176   & 0.95 & 0.03 & 0.05 & 0.03 & 0.82 & 0.03 & 0.18 & 0.03 \\
        1.250   & 0.92 & 0.03 & 0.08 & 0.03 & 0.84 & 0.03 & 0.16 & 0.03 \\
        1.333   & 0.91 & 0.03 & 0.09 & 0.03 & 0.85 & 0.03 & 0.15 & 0.03 \\
        1.538   & 0.91 & 0.04 & 0.09 & 0.04 & 0.84 & 0.04 & 0.16 & 0.04 \\
        1.818   & 0.87 & 0.05 & 0.13 & 0.05 & 0.87 & 0.05 & 0.13 & 0.05 \\
        \hline
        average & 0.93 & 0.03 & 0.07 & 0.03 & 0.84 & 0.03 & 0.16 & 0.03 \\
        \hline
    \end{tabular}
\end{table}

\section{Conclusions}

The zenith-angular dependences of particle densities $\left<\RhoMuSOO(\theta)\right>$ and $\left<\RhoSSOO(\theta)\right>$ from the total event sample with $\E \simeq 10^{19}$~eV have been analyzed by calculating mean lateral distribution functions of both components (Fig.~\ref{fig:2}). The results do not exclude that the energy estimated by Eq. \eq{eq:2} should be assumingly reduced by 10\%. This assumption requires further comprehensive analysis. The fraction of muons $\RhoMuSOO/\RhoSSOO$ in showers with zenith angles $\theta \le 38\degr$ presented in Fig.~\ref{fig:5} indicates that the cosmic ray mass composition in this energy region is close to protons. We reported this conclusion in several previous works \cite{bib:4, bib:16, bib:17, bib:18, bib:19}, where agreement between experimental data of the Yakutsk array and predictions of the \qgs{} and \qgsii{} models was noted. Some muon deficit is observed in inclined showers. The $z$ parameter \eq{eq:1} in these events is negative and becomes physically meaningless (see Table~\ref{t:1}). In our opinion, this difficulty is not due to the experimental error of the primary energy estimate, though this can- not be excluded completely for strongly inclined events. A more detailed analysis is needed for this case. The results obtained can be interpreted under the assumption of the presence of a $6-9$\% primary photons fraction in the total cosmic ray flux. We are going to continue study in this direction.

\section*{Conflict of interest}

The authors declare that they have no conflicts of interest.

\section*{Open access}

This article is licensed under a Creative Commons Attribution 4.0 International License, which permits use, sharing,
adaptation, distribution and reproduction in any medium or
format, as long as you give appropriate credit to the original
author(s) and the source, provide a link to the Creative Commons license, and indicate if changes were made. The images
or other third party material in this article are included in the
article’s Creative Commons license, unless indicated otherwise in a credit line to the material. If material is not included
in the article’s Creative Commons license and your intended
use is not permitted by statutory regulation or exceeds the
permitted use, you will need to obtain permission directly
from the copyright holder. To view a copy of this license, visit
\url{http://creativecommons.org/licenses/by/4.0/}.

\end{document}